# Lava Lakes on Io: crust age and implications for thermal output

Alessandro Mura[1], Rosaly M. C. Lopes[2], Federico Tosi[1], Peter J. Mouginis-Mark[3], Jani Radebaugh[4], Francesca Zambon[1], Matteo Paris[1], Scott Bolton[5], Alberto Adriani[1], Roberto Sordini[1], Andrea Cicchetti[1], Raffaella Noschese[1], Giuseppe Piccioni[1], Christina Plainaki[6], Giuseppe Sindoni[6]

1 Istituto Nazionale di Astrofisica – Istituto di Astrofisica e Planetologia Spaziali, Rome, Italy

2 Jet Propulsion Laboratory, California Institute of Technology, Pasadena, CA, USA

3 Hawai'i Institute Geophysics and Planetology, Honolulu, HI, USA

4 Brigham Young University, Provo, UT, USA

5 Southwest Research Institute, San Antonio, TX, USA

6 Agenzia Spaziale Italiana, Rome, Italy

Abstract: Recent observations by the JIRAM instrument onboard NASA's Juno mission have confirmed that many of Io's volcanic hot spots are active lava lakes, characterized by a colder central crust surrounded by a hotter peripheral ring. In this study, we investigate the thermal properties of thirty such lava lakes, providing new constraints on their structure and energy budget. We find that most of the total power from Io's lava lakes comes from their low-temperature crusts rather than the hotter peripheral rings, suggesting previous estimates underestimated lava lake power by up to a factor of 10. Io's paterae undergo stochastic resurfacing on timescales of roughly a decade, with each lake possibly following its own characteristic cycle. We also explore the relationship between the average temperature of the crust and the evolutionary state of each lake, offering insights into the frequency of resurfacing processes. Finally, we propose an improved assessment of Io's global thermal output, emphasizing that only full-surface observation of Io with sufficient spatial and spectral resolution can yield realistic values for the moon's volcanic total heat flux.

Key points

- Most of the total power from Io's lava lakes comes from their low-temperature crusts and not the higher-temperature border, suggesting that previous estimates of lava lakes' total power were too low by up to a factor of 10.

- A large fraction of Io's thermal output can come from faint hot spots that are lava lakes with a cold crust but with a considerable surface area.
- A resurfacing timescale for the lava lake crusts of about 10 years would explain the distribution of the observed crust temperature.
- Using the Juno JIRAM M band as a proxy for the total power of a lava lake has very large uncertainties that can be improved by using the lava lake surface area as a constraint.

# 1  Introduction

Io, Jupiter's innermost Galilean moon, hosts the most intense volcanic activity in the Solar System, powered by tidal heating from Jupiter's gravity (Peale et al., 1979). As a result, and as shown by the Voyager and Galileo missions, the surface is dominated by extensive lava flows and over 400 volcanic depressions, or paterae, such as Loki Patera, which exceeds 200 km in diameter (Carr et al., 1998; Radebaugh et al., 2001; Williams et al., 2011a). Many of Io's hot spots are concentrated within these paterae, whose thermal emissions indicate persistent, high-temperature activity consistent with lava lakes (Lopes et al., 1999; Rathbun et al., 2002; Lopes et al., 2004; Mura et al., 2024a).

To date, over 250 active centers have been identified on Io (Lopes and Spencer, 2007; Veeder et al., 2015; Zambon et al., 2023; de Kleer and Rathbun, 2023, Perry et al., 2025), many of which are lava lakes (Lopes et al. 2004), with inferred basaltic to ultramafic compositions (McEwen et al., 1998; Allen et al., 2013; Keszthelyi and Suer, 2023). Lava lakes exhibit evolving crusts formed by cooling and convection, alternating between "chaotic" and "organized" surface regimes depending on crust dynamics (Flynn et al., 1993; Lev et al., 2019; Radebaugh et al., 2016a).

Recent flybys of the Juno spacecraft in 2022 - 2025 have refined our view of Io's thermal patterns, revealing new details of its heat flow and subsurface activity (Mura et al., 2020; Zambon et al., 2023; Mura et al., 2024b, 2024a), essential for constraining models of its internal structure and volcanic energy balance. In particular, lava lakes have been observed by the Juno JIRAM (Jupiter InfraRed Auroral Mapper) instrument at M and L bands (~4.5 to 5 µm and ~3.3 to 3.6 µm, respectively) and shown to have hot rings at their margins, often complete and of constant width as can be seen at the JIRAM resolution (Mura et al. 2025a).

A fundamental issue, introduced by Mura et al. (2024a, 2025a) and discussed in more detail by Tosi et al. (2025), concerns the estimation of the total power emitted by a lava lake. In the past, several estimates of Io's total power output have been proposed without distinguishing between lava lakes, flows, and other eruptive features. Previous studies used a power-law formulation between the measured M-band power output and the inferred total output (Davies and Veeder, 2023; Perry et al., 2025). Mura et al. (2024b) used a dual-band (M/L) method for temperature estimation, as did Perry et al. (2025). In this work, we show that both approaches are largely inadequate for determining the temperature and hence the total power of a lava lake: as we demonstrate here, a lava lake typically exhibits two distinct thermal components: the hot ring (when present) and the cooler crust. Any temperature or power estimate that does not account for both inevitably leads to inaccurate results.

In this work, we present a systematic study of the total power emitted by lava lakes on Io, aiming to assess how much previous values may have been underestimated. The global power budget and the spatial distribution of thermal emission represent key constraints on Io's interior structure, as recent JIRAM observations (e.g., Pettine et al., 2024) have shown that the measured flux distribution is incompatible with purely asthenospheric heating models. Using the same analysis, we also examine the temperature distribution of lava lake crusts and, from this, attempt to infer the characteristic timescales of their resurfacing processes. The dataset is introduced in Section 2; in Section 3, we summarize the basic concepts for calculating the total power of a lava lake; Section 4 presents the results in terms of total power, and Section 5 discusses the temperature distributions and their implications for resurfacing frequency. In Section 6, we discuss the broader implications of our findings for Io's overall power budget and provide our conclusions.

## 2 Dataset

In this work, we consider most of the lava lakes studied in Mura et al. (2025a), along with one recently discovered by Juno (Mura et al., 2025c). The dataset therefore consists of images in the M band (4.5–5 µm) or L band (3.3–3.6 µm) acquired by the JIRAM instrument aboard Juno. Although originally designed for the study of Jupiter, JIRAM has also been used for observations of Io, as described in Mura et al. (2020). Of the approximately 50 features included in Mura et al. (2025a), about 30% had to be excluded from the current study. In some cases, this was due to the limited spatial resolution of the JIRAM images, which allow the detection of the hot margins but do not provide sufficient spatial resolution to study the crust temperature. In other cases, features were excluded due to their intrinsic characteristics, such as irregular emission patterns or the presence of unclear features within the crust.

Among the hot spots selected for this study, some warrant particular mention. Amaterasu Patera (see Table 1 for coordinates and time of observation and figure S1 in Mura, 2026) was observed during orbit 51 on May 16$^{th}$ 2023, showing the characteristic ring-shaped morphology with a cooler central crust; it was observed again during orbit 55 on October 15$^{th}$, 2023, this time without any evidence of a ring, but with a generally hotter surface. It can be hypothesized, although not with absolute certainty, that some resurfacing event, perhaps similar to those previously observed at Loki by de Kleer and de Pater (2017) and Mura et al. (2025b), occurred in the intervening period. This observation is therefore useful for evaluating the possible frequency of resurfacing events, although alternative explanations will be discussed below.

Another noteworthy case is Chors Patera, both because it was previously discussed in Mura et al. (2024a, 2025a) and Tosi et al. (2025), and because it was observed at the highest possible resolution in both bands (in PJ55), showing a uniform surface without ambiguity in both the L and M bands. To strengthen our analysis, we also use data for Chors Patera from PJ51, which has a lower spatial resolution but still sufficient for analysis. Table 1 lists the features selected for this study.

*Table 1. Dataset for this study. Location of hot spots, time of observation and orbit of Juno, spectral band of the observation.*

| Name | Latitude | Longitude E | PJ (Orbit) | Date | Band |
|---|---|---|---|---|---|
| Amaterasu | 38 | 53 | 51 | 2023 05 15 | M |
| Amaterasu | 38 | 53 | 55 | 2023 10 15 | M |
| Babbar | -40 | 89 | 60 | 2024 04 09 | M |
| Bulicame/P77 | 28 | 169 | 51 | 2023 05 15 | M |
| Catha | -53 | -102 | 62 | 2024 06 13 | M |
| Chors | 68 | 110 | 55 | 2023 10 15 | L |
| Chors | 68 | 110 | 51 | 2023 05 15 | M |
| Chors | 68 | 110 | 55 | 2023 10 15 | M |
| Dazhbog | 55 | 57 | 51 | 2023 05 15 | M |
| Dusura | 38 | -119 | 55 | 2023 10 15 | M |
| Fuchi | 29 | 32 | 55 | 2023 10 15 | M |
| Heno | -57 | 48 | 60 | 2024 04 09 | M |
| Kinich Ahau A | 49 | 50 | 55 | 2023 10 15 | M |
| Lei-Kung C | 72 | 87 | 55 | 2023 10 15 | M |
| Lei-Kung K | 75 | 104 | 55 | 2023 10 15 | M |
| Maui | 17 | -124 | 55 | 2023 10 15 | M |
| Mazda A | -9 | 47 | 60 | 2024 04 09 | M |
| Mihr | -16 | 55 | 55 | 2023 10 15 | M |
| Mulungu | 17 | 142 | 51 | 2023 05 15 | M |
| P139 | -62 | 162 | 68 | 2024 12 27 | M |
| P62/E. Surya | 22 | -146 | 55 | 2023 10 15 | M |
| P63 | 35 | -138 | 55 | 2023 10 15 | M |
| P63 | 35 | -138 | 55 | 2023 10 15 | L |
| PFd1069/PFd1293 | 44 | 12 | 55 | 2023 10 15 | M |
| PFd630/Pajonn | 83 | -89 | 55 | 2023 10 15 | M |
| PFu1128 | 30 | 130 | 51 | 2023 05 15 | M |
| Pfu1063 | 44 | 3 | 57 | 2023 12 30 | M |
| Sigurd | -6 | -98 | 55 | 2023 10 15 | L |
| Surt | 45 | 23 | 55 | 2023 10 15 | M |
| Susanoo | 22 | 140 | 51 | 2023 05 15 | M |
| Tol Ava | 2 | 37 | 58 | 2024 02 03 | M |
| Tvashtar A | 63 | -126 | 55 | 2023 10 15 | M |
| Unnamed 07 | 52 | -63 | 57 | 2023 12 30 | M |
| Unnamed 08 | 52 | -29 | 57 | 2023 12 30 | M |
| Unnamed 18/JRM048 | -83 | -66 | 62 | 2024 06 13 | M |
| Uta | -35 | -22 | 58 | 2024 02 03 | M |

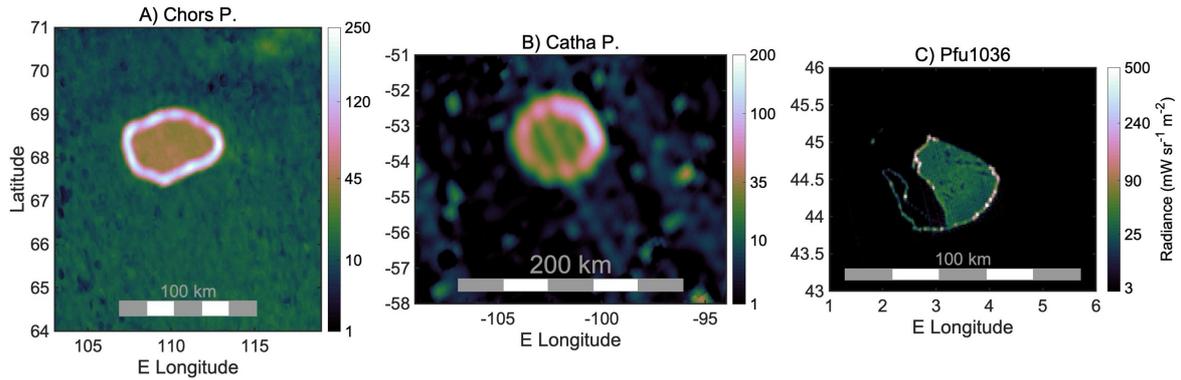

*Figure 1: three examples of lava lakes with clear evidence of an outer hot ring and a colder, yet emitting, inner crust. Panel A: Chors Patera in orbit 55; Panel B: Catha Patera in orbit 62; Panel C: Pfu1036 in orbit 57. Colors are coded according to M-band radiances, all units in mW sr$^{-1}$ m$^{-2}$. Adapted from Mura et al., 2025a, and Tosi et al., 2025*

## 3 Model

A calculation of the power output of lava lakes for some representative cases was performed by Tosi et al. (2025). In that paper, the cases of Chors Patera, PFu1036 and Catha Patera were discussed. These three features are shown in Figure 1. In this section we introduce the main concepts of that analysis by using another example, based on a feature known as P63 (see Figure 2). Then we apply the same calculation more broadly, to show how frequently this paradigm (a lava lake with hotter, brighter margins but an interior crust that actually emits most of the power) is found on Io. We use most of the features observed by Mura et al. (2024a), when the quality of data permits it, plus one observed in Mura et al. (2025c).

JIRAM observed P63 (Figures S22 and S23 in Mura, 2026) on PJ55 (2023/10/15), and found that the patera emitted, at that time, about ~0.4 GW of power ($P_M$) in the M band (~4.5 to 5 µm, 0.5 µm passband), from both the ring and the crust (but mainly from the hotter ring); in the L band (~3.3 to 3.6 µm, 0.3 passband), the total power is 0.15 GW. In the M band, the inner crust exhibits a radiance of 40 mW sr$^{-1}$ m$^{-2}$, in the L band the inner crust is basically below the minimum detectable signal. This ensures, beyond any doubt, that what we measure on the crust in the M band is genuine emission and not stray light from the surrounding ring; if the latter were true, we should see more crustal emission in the L-band. Another key quantity for the following calculation is based on estimating the area (~500 km²) of P63 based on the visible image (see Methods). Figure 2 shows visible, M-band and L-band maps of P63. The area is calculated as the extension of the visible-range dark ring and its interior (corresponding to recent lava; Figure 2, panel A). Using this area, the emission from the crust can be estimated. In fact, the measured band radiance (~0.05 W/m²/sr) corresponds to a brightness temperature of ~230 K, implying a radiant emittance of ~160 W m$^{-2}$ if integrated over all wavelengths (as explained in the Appendix, the key point is that brightness temperature is meaningful when there are no sub-pixel structures; at the spatial resolution used here, it therefore provides a reliable estimate of the crust temperature). Scaled to the full surface area, this gives ~80 GW of total emission ($P_T$). The median value of the measurements of the total power obtained by using the M/L band ratio, from Perry et al. (2025) is 7 GW, that is, an underestimation of a factor of 10 under our assumptions. Incidentally, 7 GW is also the result of our analysis when combining L and M band radiances integrated over both the ring and the crust (respectively 0.15 and 0.4 GW, yielding a temperature of 550 K, which is somewhat an intermediate temperature between the crust and the hot ring). Even when adopting the relationship from Davies and Veeder (2023) (also used in Perry et al., 2025, Eq. D1) $y = 23.702\, x^{0.8838}$, where $y$ is hot spot total thermal emission in GW, and $x$ is 4.8-μm spectral power in GW μm$^{-1}$, also used in Perry et al. (2025, Eq. D1), starting from an M-band total power of 0.4 GW yields an estimated total emission of only ~20 GW, which is still underestimated by about a factor of four.

This discrepancy highlights the extreme sensitivity of power estimates to temperature, and the difficulty of deriving robust values from M-band measurements alone.

Tosi et al. (2025) explained the origin of the problem. The temperature distribution has two very different peaks: one for the ring (for the surrounding lava ring, Mura et al., 2024a adopted a temperature of 900 K) and a much colder one for the crust (as we have seen, about 230 K in this case). Trying to fit a single temperature value, combining L and M, gives 550 K, which does not have much physical meaning. Using it as a representative temperature to scale the M-band power yields, as mentioned, 7 GW—less than one-tenth of the actual value.

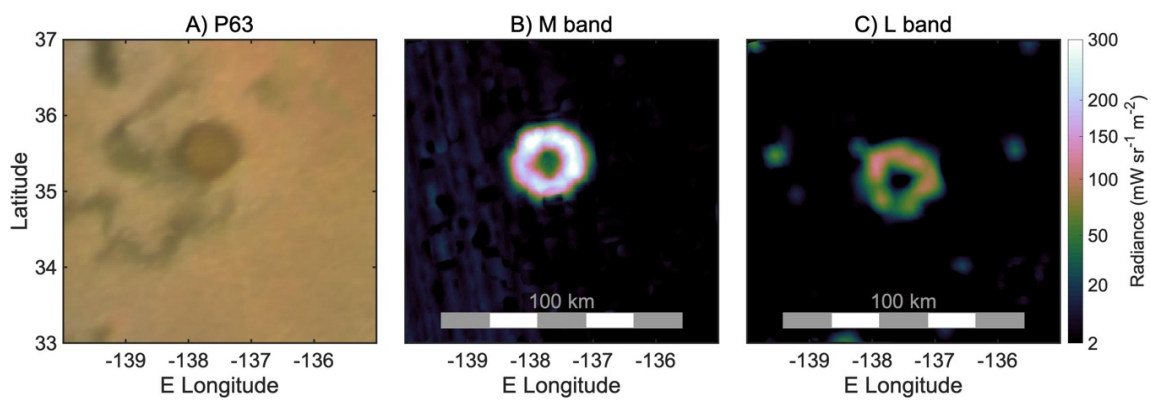

Figure 2. Feature named P63 in the visible (panel A), M band (panel B) and L band (panel C), observed in orbit 55. **Visible data is from Williams et al., (2011a, 2011b).**

Figures presented in Mura et al. (2024a) illustrate this point clearly: only a small fraction of the total emission is captured in the M-band, and the spectral power (i.e. the spectral radiance integrated over the solid angle and the area of the lake) comparison between crust and ring emphasizes the dominance of longer wavelengths on the total thermal output. Figure 3 shows the spectral power (spectral radiance integrated over solid angle and area) coming from the hotter ring of P63 and the spectral power coming from the colder, but more areally extensive, crust.

An alternative scenario to explain the thermal pattern is that the lake surface consists of a much colder crust, pervaded by very hot sub-pixel fractures. Terrestrial lava lakes commonly display fractures with exposed lava both at the margins and within the central crust (e.g., Radebaugh et al., 2016b), which could suggest a similar scenario on Io. For this scenario to hold, such hypothetical sub-pixel fractures would need to be perfectly uniformly distributed across the lake interior. Sub-pixel fractures down to 3 m (Mura et al., 2024a) are indeed observed at the lake margins but show no detectable signature in the center, demonstrating that JIRAM is capable of detecting such features when their radiance contrast is sufficient. Conversely, central fractures are absent in JIRAM data of Ionian lakes. Moreover, in many cases the crust temperatures are already very low (~200 K), implying long cooling times (>10 y) and substantial crust thickness (>10 m). For the crust to contribute nothing to the thermal output, leaving only sparsely distributed hypothetical sub-pixel fractures to contribute to the observed radiance, the temperature of the crust would need to be even lower, near Io's background surface, which is implausible given the decades-long activity of these volcanic centers.

However, the most important objection to this scenario is the complete absence of signal in the L band coming from the center of the lake. If the radiance in the center of the lake is coming from several sub-pixel hot fractures in a very cold crust, the signal in the L band should be comparable to that in the M band (see Figure 6 of the supplementary material from Mura et al., 2024b, on the L/M ratio), but we do not see an L-band signature. Therefore, the hypothesis of a cold crust pervaded by unresolved sub-pixel hot fractures is inconsistent with the observations.

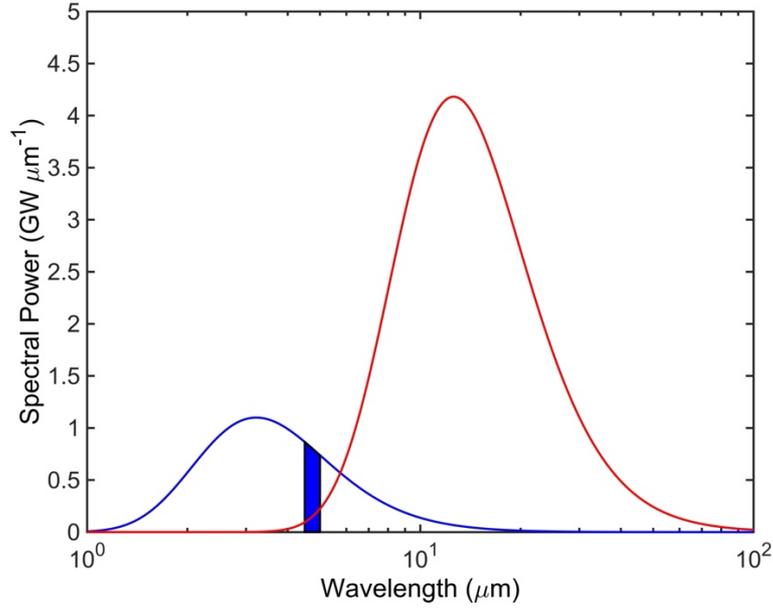

*Figure 3: in blue, the spectral power (spectral radiance integrated over solid angle and area) coming from the hotter ring of P63 (at 900 K); in red, the spectral power coming from the colder, but much larger, crust (at 230 K).* **The area for the crust is ~500 km² (see main text); the area for the hotter ring is inferred from the measured M-band power as in Mura et al. (2024a) and Tosi et al. (2025).** *The blue area represents the integrated signal captured in the M band of JIRAM (or any other instrument working in the 4.5-5 µm range). Adapted from Tosi et al. (2025) to the case of P63.*

There are also slightly different cases, wherein we are not dealing with a lava lake but with a different kind of hot spot, which does not have a hot ring. If such hot spot is observed with insufficient spatial resolution, as may happen for ground-based observations, it may be impossible to extract surface areas of the hot spot independently. In these cases, the brightness temperature does not provide any information about the actual temperature, because what is measured is not the true radiance, but the radiance multiplied by the pixel's filling factor. In some cases, it may be unclear whether a hot spot is a lava lake or a different thermal feature, such as a recently emplaced lava flow. For this reason, we include in our analysis hot spots like Tvashtar, which is currently not a lava lake, in order to assess how the total emitted power may be underestimated when such features are observed at lower spatial resolution than in the present study. As we will show, also in the case of such hot spots, using the M-band radiance only leads to underestimating the total power.

In the next section, we will replicate this analysis (where possible) for all lava lakes observed by JIRAM. The full description of the method is given in Appendix 1.

## 4 Total thermal output from lava lakes

After performing the calculation described in the previous section for several lava lakes where the spatial resolution and data quality were sufficient for this purpose, we present the results in this section. Figure 4 shows a global map of the locations of the investigated hotspots; IR radiance maps of single hot spots are given in Mura (2025), and referred as Figures S1 to S36. Table 2 shows the results of our calculations for the 32 paterae (some observed more than once) in our study. $P_M$ is the total power in the M band as directly measured by JIRAM. $P_T$ is the total power integrated over all wavelengths, and calculated as described in section 3. The temperature, area and crust age are calculated as described in the Appendix. Of the 32 paterae studied, 5 were uncertain because the crust radiance was too low to measure (these are noted "1" in Table 2) and 3 were uncertain for other reasons (noted "3" in Table 2). All these 8 are noted as red in Table 2 (and in Figure 5, see below) because of these uncertainties. In the following, we offer some additional remarks on these entries in Table 2, mainly to provide further context.

To begin with, we have deliberately included a few lava lakes whose interiors appear completely cold, even in the M band. By "cold" we mean that their emission is below the noise level, bearing in mind that the background radiance outside the patera has been subtracted. We retain these cases because, although their power estimates are rather uncertain (they are flagged in red in Figure 5), they are nonetheless indicative of a cold crust and are in some way useful for analyzing the distribution of crustal ages within a lake. Bulicame/P77 (Fig. S4), Lei-Kung K (Figs S14 and S15), Mazda A (Fig. S17), PFd1069 (or PFd1293; the naming convention remains uncertain, Fig. S24), and Unnamed 07 (Fig. S33) belong to this group. The estimation of the crust age is based on the model of crust cooling from de Kleer and de Pater (2017) and it is discussed in the appendix.

In particular, Unnamed 07 exhibits an extremely well-defined hot ring, but it is covered by JIRAM only for about half of its extent, and the crust is too cold (T<190) for a reliable temperature determination. In all these cases we have arbitrarily indicated the temperature as N/A. Since we are nevertheless confident that the crust is cold and therefore sufficiently old (see panel A of Figure 1 in de Kleer and de Pater, 2017, a model also supported by Mura et al., 2025b), we have assumed a crustal age > 10 years. This arbitrary timescale can easily be changed in subsequent calculations; thus, it is not crucial to further discuss this assumption at this stage. In some of these cases, however, we have provided the best possible estimate of the total power, while emphasizing for the sake of accuracy that these are uncertain cases, marked as red points in Figure 5.

We then have the cases where a robust and reliable estimate of the crustal radiance is available: Amaterasu (Figs S1 and S2), Babbar (Fig. S3), Catha (Fig. 1), Chors (Figs. 1, S6, S7, and S8), Dazhbog (Fig. S9), Dusura (Fig. S10), Fuchi (Fig. S11) showing a weak signal but consistent across different radiance estimation methods; see "Methods"), Lei-Kung C (Fig. S15), Maui (Fig. S16), Mihr (Fig. S19, also a weak but coherent signal), Mulungu (Fig. S19), P139 (Fig. S20), P62 (also known as E. Surya, Fig. S21), P63 (Fig. S22), Pajonn (Fig. S25, previously known as PFd630), PFu1128 (Fig. S26), Pfu1063 (Fig. S27), Surt (Fig. S29), Tol Ava (Fig. S31), Tvashtar A (Fig. S32), Unnamed 08 (Fig. S34), Unnamed 18/JRM048 (Fig. S35), and Uta (Fig.S36). Among these, it is worth noting that Amaterasu was observed in two different perijove (PJ) passes: during PJ51 it exhibited the typical appearance of a lava lake with a hot ring, whereas in PJ55 it showed a more uniform morphology at comparable spatial resolutions. Presenting two different observations of the same lava lake is intentional, as it provides complementary information on different possible stages of a lava lake at different times. Similarly, for Chors we present two cases (PJ51 and PJ55), and we also include an estimate obtained in the L band (not shown in Figure 5). For P63 we show both L and M bands (see Figure 2), whereas for Sigurd (Fig. S28) only an L-band estimate exists, which we report solely for completeness.

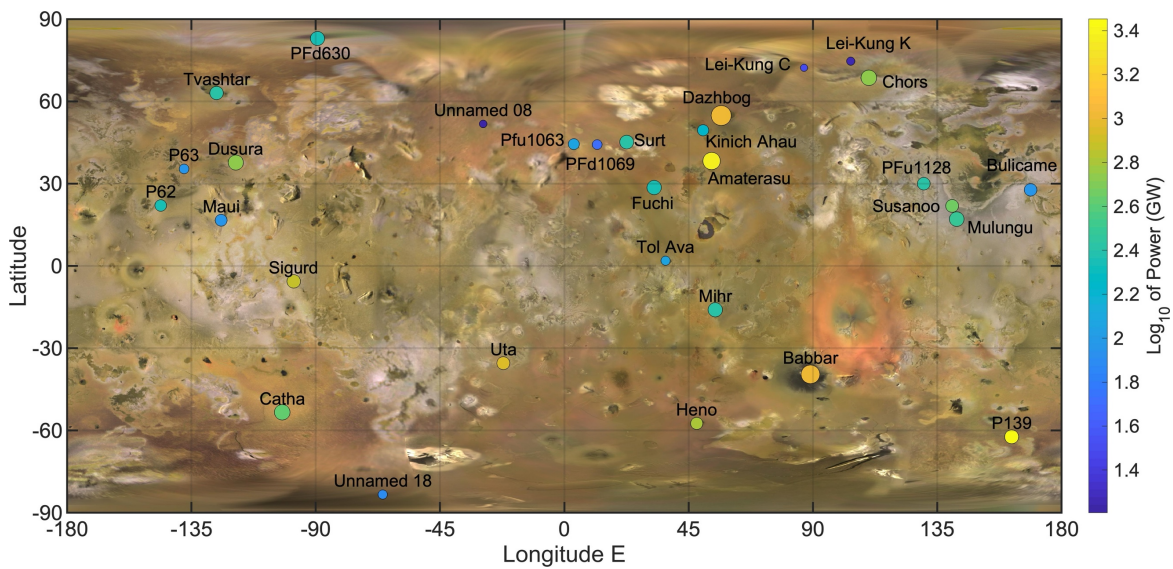

*Figure 4: map of the hot spots investigated in this work. Size of circles is roughly proportional to the hot spot area; color is coded according to total power output.*

Heno (Fig. S12) represents a somewhat complex case; we flag it in red because the patera is evidently active over only about one quarter of its total surface, and the hot ring is highly irregular. Kinich Ahau (Fig. S13) is very hot even at its center, with only a vaguely discernible ring-like structure. The absence of a clear ring does not in principle compromise the analysis, provided that the minimum temperature can be estimated with sufficient confidence; therefore, we retain it in the table but mark it in red in Figure 5. P139 (Fig. S20) appears to be a lava lake that has recently resumed activity (see Mura et al., 2025a), while P62 (Fig. S21) is in a phase where no surrounding ring is visible. However, its infrared morphology is highly consistent with its optical counterpart, so there is no risk of overestimating the active area, and it can be safely included. The same applies to Pajoon (PFd630), which has very good spatial resolution and uniform radiance (although in this case the visible counterpart provides no additional information, as no high-resolution optical images are available). Pfu1063 (Fig. S27) is active only over part of its area, and only that portion has been considered. No visible images are available, and the area was estimated from the infrared data. Tol Ava shows uniform radiance and no distinct ring, but the agreement between visible and infrared observations allows the crustal temperature to be reliably estimated. Tvashtar corresponds to a lava flow (Lopes et al., 2025), and it is included to extend the analysis to this class of Io features.

Among the other cases that must be treated with caution and have been excluded from the analysis—but are reported here for completeness—is Susanoo (Fig. S30), whose radiance exhibits an asymmetric morphology between north and south. We include it and provide a preliminary estimate only because the edge of the IR image matches very well the corresponding visible outline.

| M-band observations Name | PJ | $P_M$ GW ±10% | $P_T$ GW | T K ±5% | Crust Age Years ±25% | Area km² | Area (USGS) km² | $P_T$ (literature) GW | Notes | Figure ref. |
|---|---|---|---|---|---|---|---|---|---|---|
| Amaterasu | 51 | 41 | 2400 | 290 | 0.67 | 5300 | 7100 | 1300 | | S1 |
| Amaterasu | 55 | 14 | 2300 | 300 | 0.45 | 5300 | 7100 | 1300 | | S2 |
| Babbar | 60 | 1.7 | 1000 | 220 | 5.1 | 7200 | 9600 | 47 | | S3 |
| Bulicame/P77 | 51 | 0.36 | 91 | N/A | >10 | 1500 | N/A | 9.2 | (1) | S4 |
| Catha | 62 | 0.73 | 430 | 220 | 6.3 | 3400 | 3400 | 7.7 | | S5 |
| Chors | 51 | 1.1 | 540 | 230 | 3.8 | 3400 | 3700 | 28 | | S7 |
| Chors | 55 | 0.67 | 460 | 220 | 5.5 | 3400 | 3700 | 28 | | S8 |
| Dazhbog | 51 | 1.3 | 1000 | 210 | 8.6 | 9400 | 11000 | 58 | (2) | S9 |
| Dusura | 55 | 0.8 | 540 | 240 | 2.9 | 3100 | 3400 | 310 | | S10 |
| Fuchi | 55 | 1.2 | 220 | 200 | 14 | 2500 | 3500 | 36 | | S11 |
| Heno | 60 | 4.2 | 630 | 300 | 0.41 | 1300 | N/A | 26 | (3) | S12 |
| Kinich Ahau A | 55 | 0.49 | 170 | 240 | 3 | 950 | N/A | 41 | (3) | S13 |
| Lei-Kung C | 55 | 0.17 | 31 | 240 | 2.9 | 160 | N/A | 6 | | S14 |
| Lei-Kung K | 55 | 0.035 | 18 | N/A | >10 | 270 | N/A | 0.7 | (1) | S15 |
| Maui | 55 | 0.23 | 93 | 190 | 16 | 1200 | 1100 | 4.9 | | S16 |
| Mazda A | 60 | 0.23 | N/A | N/A | >10 | 890 | 39000 | 19 | (1) | S17 |
| Mihr | 55 | 0.47 | 270 | 210 | 9 | 2500 | 2700 | 12 | | S18 |
| Mulungu | 51 | 0.54 | 320 | 210 | 7.6 | 2700 | 3300 | 30 | | S19 |
| P139 | 68 | 82 | 2800 | 350 | 0.12 | 2200 | N/A | N/A | (4) | S20 |
| P62/E. Surya | 55 | 0.39 | 220 | 250 | 2.2 | 1100 | N/A | 130 | | 21 |
| P63 | 55 | 0.44 | 88 | 230 | 3.4 | 520 | N/A | 7 | | 22 |
| PFd1069/PFd1293 | 55 | 0.067 | 53 | N/A | >10 | 550 | N/A | 2.9 | (1) | 24 |
| Pajonn/P Fd630 | 55 | 0.081 | 230 | 210 | 11 | 2300 | N/A | 120 | | 25 |
| PFu1128 | 51 | 1 | 260 | 240 | 3.2 | 1500 | N/A | 15 | | 26 |
| Pfu1063 | 57 | 0.13 | 110 | 220 | 5.6 | 860 | N/A | 13 | | 27 |
| Surt | 55 | 0.8 | 260 | 220 | 7.3 | 2100 | 2200 | 29 | | 29 |
| Susanoo | 51 | 2.5 | 460 | 270 | 1 | 1600 | 2400 | 200 | (3) | 30 |
| Tol Ava | 58 | 0.28 | 120 | 270 | 1.2 | 430 | 6000 | 130 | | 31 |
| Tvashtar A | 55 | 0.28 | 260 | 230 | 5 | 1900 | N/A | 84 | | 32 |
| Unnamed 07 | 57 | 0.034 | 0 | N/A | >10 | 230 | N/A | N/A | (1) | 33 |
| Unnamed 08 | 57 | 0.0066 | 16 | 190 | 16 | 210 | N/A | N/A | | 34 |
| Unnamed 18/JRM048 | 62 | 0.29 | 75 | 230 | 3.2 | 440 | N/A | 830 | | 35 |
| Uta | 58 | 11 | 880 | 320 | 0.29 | 1600 | 900 | 130 | (2) | 36 |
| L-band observations | | $P_L$ GW | | | | | | | | |
| Chors | 55 | 0.14 | 1.8 | N/A | N/A | 3400 | 3700 | 28 | (5) | S6 |
| P63 | 55 | 0.16 | 130 | 260 | 1.4 | 520 | N/A | 7 | (5) | S23 |
| Sigurd | 55 | 0.33 | 740 | 290 | 0.58 | 1900 | 2200 | 920 | (5) | S28 |

*Table 2: Data for the hot spots analyzed in this study. Columns are: Juno's orbit for the observation (PJ); total power in the M band ($P_M$), or in the L band ($P_L$, last three rows); total power ($P_T$); crust temperature (T) and estimated age; area of the patera; total power from literature, from Perry et al. (2025); the last column indicates the number of the figure in the catalogue of Mura (2025). Notes are: (1) crust radiance and temperature are too low, the power estimate is not reliable, and the temperature is not given intentionally; used only for crust age analysis, and shown as red dots in Figure 5; (2) these cases use method "2" for crust temperature estimation (see Appendix); (3) these cases are suspect or peculiar in some way, they are not used for crust age analysis, and shown as red dots in Figure 5; (4) the very recent outburst in the Illyrikon Regio; (5) L-band data, shown just for completeness, not used for Figure 5 or for crust age analysis. Uncertainties are given globally and are, on average, 10% for the total power estimation, 5% for the crust temperature estimation and at least 25% for the crust age. The uncertainty on the total power ($\Delta P_T$) does not include the uncertainty on the area, which is not known, and hence $\Delta P_T$ may be underestimated.*

Figure 5 presents the experimental data from this study, plotting the spectral power in the M-band (centered at approximately 4.78 µm, and with 0.5 µm of bandwidth) in GW/µm on the x-axis against the total power in GW on the *y*-axis. For comparison, a dashed line indicates the empirical relationship employed in previous studies (Davies and Veeder, 2023; Perry et al., 2025).

The analysis of Figure 5 reveals several key findings. First, a (weak) correlation exists between the M-band spectral power and the total power, which is an expected outcome. Larger lava lakes, possessing a greater perimeter, should inherently exhibit a more extensive hot boundary and thus emit more power across both the crust and the margins. Second, and more critically, the empirical curve from previous works (Davies and Veeder, 2023, Perry et al., 2025, dashed line, see figure caption) falls entirely outside our population, getting close only with the single data point at the highest spectral power (which was P139, a hot spot detected during the largest eruption recorded so far at Io (Mura et al., 2025c), hence a recently active hot spot and admittedly not a very representative one). This discrepancy demonstrates that earlier estimates are likely subject to a very large systematic error, particularly at low-to-medium M-band power levels. In fact, this relationship (used by Perry et al., 2025, and proposed by Davies and Veeder, 2023) connects total thermal emission to spectral power at 4.8 µm, and it is statistically robust for the temperature range of 400-600 K; however, its application becomes misleading outside this temperature range, or when hot spots have a structure with two distinct temperature components. Finally, even adjusting this power-law fit to our experimental data points, the average discrepancy between fitted and true values is about a factor of two. This shows that, although a correlation is present, the M-band alone remains an unreliable proxy for total power. We conclude that it is highly probable that prior analyses resulted in a significant statistical underestimation of an important component of the total power budget, affecting both the derived total power and its inferred spatial distribution. This may help explain why previous studies on the distribution of volcanic power on Io have yielded conflicting results (see, for instance, Pettine et al. 2024; Davies et al. 2024, and the comparison with gravity measurements in Park et al., 2025).

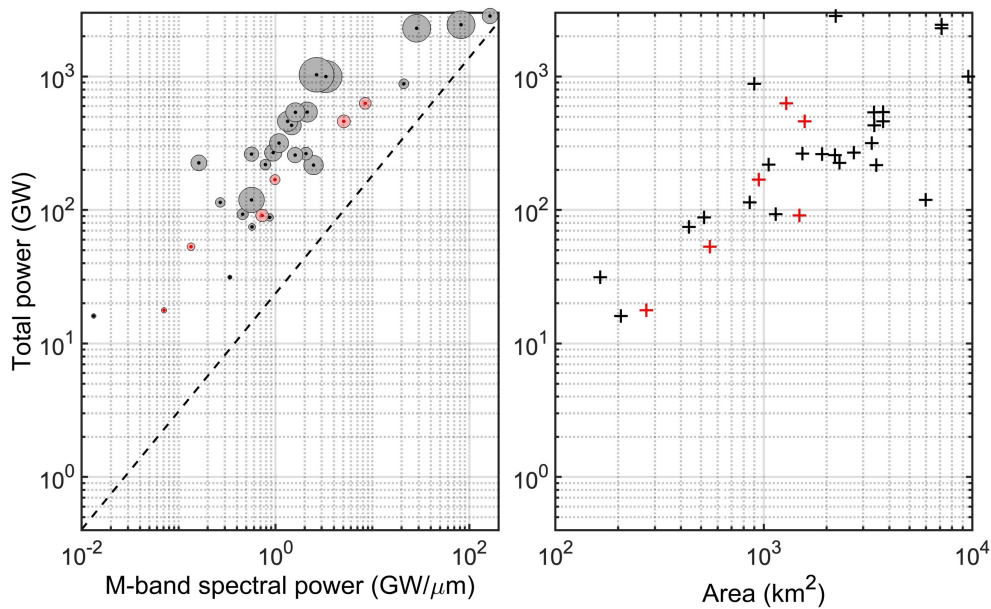

*Figure 5: Left: gray circles represent the spectral power in the M-band (centered at approximately 4.78 µm, and with 0.5 µm of bandwidth) in GW/µm on the x axis plotted against the total power in GW on the y axis. For comparison, a dashed curve indicates the empirical relationship employed in Davies and Veeder (2023) and Perry et al. (2025): $y = 23.702x^{0.8838}$, where y is the total thermal emission in GW and x is the 4.8-µm spectral power in GW/µm. Circles are sized according to the area of the patera/lake. Red circles/dots are cases where the estimation is less reliable (see text and Table 2 for details). Right: same, plotting the total power in GW on the y axis as a function of the area of the patera/lake on the x axis.*

We also recall another potential source of error, already noted in Mura et al. (2024a): for emission angles different from 0°, the rim of the patera partially obscures the underlying lava lake, since the lake surface always lies below the surrounding terrain. At high emission angles and without sufficient spatial resolution to discriminate the crust from the ring — which, for example, often occur for non-equatorial paterae observed from Earth — this effect can lead to a further underestimation of the observed power.

It is noteworthy that we can also generate a scatter plot of lava lake surface area versus total power (Figure 5, right panel). In this case, since the surface area is a direct geometric factor used in the calculation of total power, the dispersion of data points around a perfect correlation is primarily attributable to the uncertainty in the crust's surface temperature. The observed correlation in this area-versus-power plot is not significantly worse than that of the M-band spectral power versus total power (0.6 vs. 0.8). The best-fitting relation in this case is obviously *y=c \* area*. This leads to the seemingly paradoxical conclusion that the simple surface area may serve as a proxy for total power with a reliability comparable to that of an actual infrared M-band measurement. While this is obviously a method with limited reliability, it is plausible that combining these two sources of information —M-band flux and surface area—could yield a more accurate estimate of total power. However, we leave such a multivariate analysis for future work, as it falls beyond the scope of the present study.

## 5 Crust temperature and resurfacing timescales

Figure 6 shows the temperature distribution derived from the data in Table 2, represented as a histogram with regular bins between 200 and 350 K, together with a kernel-smoothed density estimation ("ks-density", dashed line in Figure 6, see caption), which exhibits a peak around 220–225 K. The maximum of the distribution lies in this range, followed by a long but sparsely populated tail extending up to 350 K. This tail corresponds to case P139, a rather uncommon example among the observed lava lakes, since the most recent resurfacing event is estimated to have occurred only one or two months earlier, as the same patera appeared completely dark in the Juno observation preceding the one in which it was seen glowing.

Amaterasu was observed at two times in 2023 in two distinct stages of its activity: an "annular" phase and another apparently following a resurfacing episode, characterized by a slightly warmer crust and the absence of the annular feature. However, the case of Amaterasu may not necessarily correspond to a resurfacing event; it could instead reflect an observation taken during a quiescent phase, when the hypothesized piston-like up-and-down motion (Mura et al., 2024a) was momentarily halted—possibly at a maximum or minimum level. The same reasoning could apply to other cases lacking an annular pattern. In such instances, the appearance of a slightly warmer crust might simply result from measurement uncertainties.

Chors was observed at two different epochs. This serves both to improve statistical robustness and to provide a temperature comparison over a time span of about six months. The two measurements are fairly consistent: the second temperature is about 10 K lower than the first. Although this difference does not exactly match the expected evolution over a six-month interval, it comfortably falls within the estimated uncertainty, which we recall being at least 25% for the crust age.

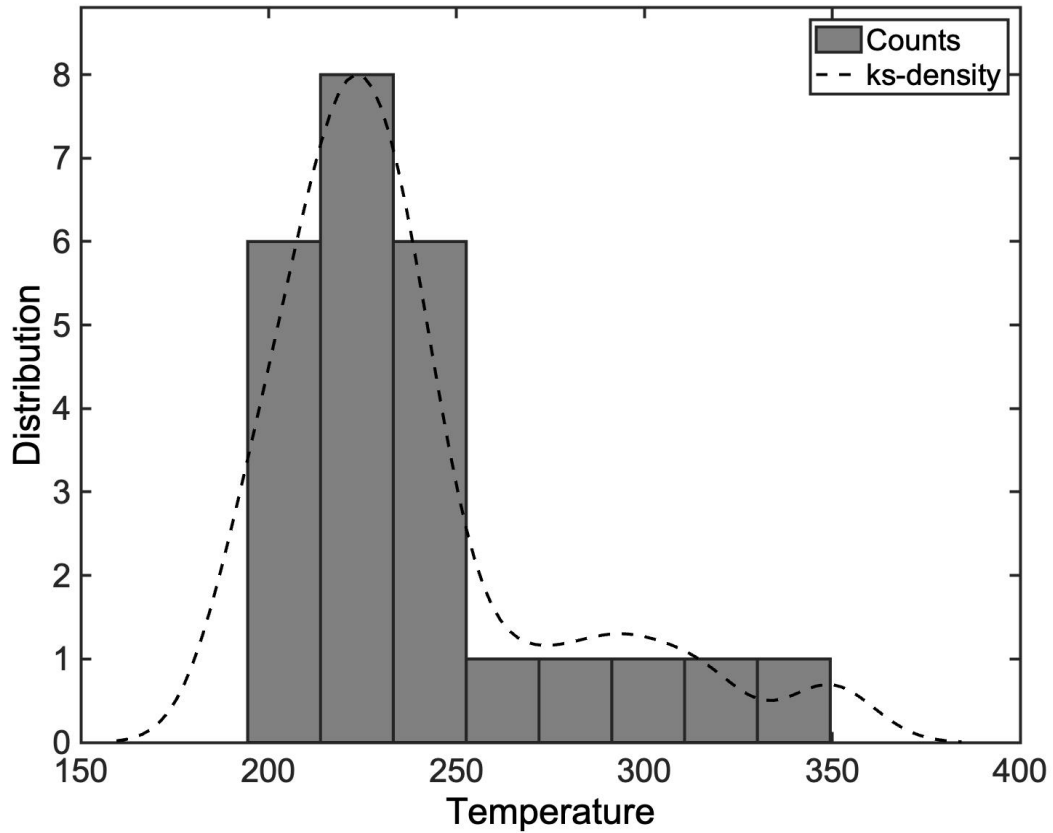

*Figure 6: distribution of inferred crust temperature from Table 2.* ***The dashed line represents a kernel density estimation ("ks-density") of the temperature distribution. A Gaussian kernel was used. This provides a smooth estimate of the underlying probability density of the temperatures.***

From the measured temperature, it is possible to infer the age of the crust, following the analysis presented in de Kleer and de Pater (2017) (panel A of their Figure 1). The model predicts, for instance, an age of approximately 13 years for a 200 K crust, about 2 years for 250 K, roughly 5 months for 300 K, roughly 1 month for 350 K, allowing interpolation for intermediate values. We assume that this estimated age corresponds to the time elapsed since the last resurfacing event up to the observation epoch. As detailed in the Methods section, we further assume that resurfacing events in lava lakes follow a Poisson statistical distribution (Burgi et al., 2002; example of a Poisson-based statistical approach for terrestrial lava lakes). Based on this assumption, we computed and maximized the likelihood function, obtaining a characteristic timescale $\tau$ for resurfacing events of about 8 years (when assigning an age >10 years to "too cold" paterae). If this lower limit is instead set to >20 years, the resulting characteristic timescale slightly increases (to $\tau \approx 10$ years).

Finally, a Kolmogorov–Smirnov test ($p = 0.35$) does not reject the null hypothesis that the crust ages follow an exponential distribution within the available statistics (see methods for details), supporting the assumption of a resurfacing process with Poisson temporal distribution. One could in principle consider performing a formal test to check whether the observed crust ages are compatible with the hypothesis that each lava lake undergoes resurfacing at a fixed period specific to that lake, potentially different from all other lakes — essentially treating each lake as an individual "Loki-type" patera (Rathbun et al., 2002). However, since we do not have time-resolved observations of resurfacing for each lake, it is impossible to perform such a test or to evaluate whether this scenario is valid. Therefore, the hypothesis that each lake has its own resurfacing cycle cannot be meaningfully confirmed or refuted with the available data.

In summary, the distribution of measured crustal ages rules out the - very implausible - possibility that all lava lakes follow the same resurfacing period (see appendix). They may still undergo periodic resurfacing—each with its own characteristic timescale—or experience stochastic events and distinguishing between these scenarios remains challenging. During resurfacing events, lava may spill over the top of the crust (e.g. Howell et al., 2014), though crust foundering has also been proposed (Rathbun et al., 2002). One particularly informative case provides useful clues: Amaterasu Patera was observed twice, five months apart. On 15 May 2023, the crustal age was estimated at roughly eight months, whereas on 15 October 2023—five months later—the inferred age was about five months. The October estimate aligns with resurfacing around May, shortly after the first observation, and the May value remains compatible with a ~5-month periodicity, considering the ≥25% uncertainty.. Cycles shorter than one year are plausible for Amaterasu, while other lakes show longer periods due to colder, older crusts. Though a single example, Amaterasu suggests the idea that each lava lake may follow its own intrinsic periodicity—a scenario strengthened by Loki, which is known to exhibit quasi-periodic behavior on longer timescales (and at much larger spatial scales, Rathbun et al., 2002). What complicates and, in a sense, challenges this analysis, however, is the fact that in the two observations by Amaterasu, one shows the hot ring while the other does not, which can only be explained by a halt in the hypothesized piston-like movement (Mura et al., 2024a). A mere change in observation geometry, as discussed in the same study, would not reproduce the effect observed in the Amaterasu data. This, therefore, demonstrates how the lava lake environment on Io is dynamic and difficult to analyze.

In conclusion, our results support the hypothesis that Io's paterae undergo stochastic or periodic resurfacing events on typical timescales of roughly a decade, and a compelling possibility, to be further investigated, is that each system maintains its own more-or-less regular resurfacing cycle, with characteristic periods that differ from one lake to another.

## 6 Discussion and conclusions

This work focuses on the analysis of thirty volcanic hot spots on Io, most of which are lava lakes, alongside other surface features. We measured the crustal temperatures of these sites and used these data to conduct two complementary studies. The first assesses Io's total thermal power output, revealing that previous estimates were significantly underestimated. The second investigates the evolutionary state of the lava lake crusts, providing constraints on their resurfacing timescales and the distribution of observed temperatures.

### 6.1 Total power output.

Figure 5 clearly shows the dramatic underestimation that can occur in the total power of lava lakes when using a function based solely on the spectral power in the M band to estimate the total output. One might wonder whether incorporating additional information from the L band (3.45 μm), for example, could improve the situation. We can anticipate that the answer is no. As seen in Figure 3, the real issue for lava lakes is the lack of spectral information from about 5 μm up to 20 μm and beyond, not the precision of our measurements below 5 μm.

Nevertheless, to substantiate this argument, we compared our results in Table 2 with the most recent estimates based on the L/M ratio published by Perry et al. (2025), where available, and averaging over all JIRAM measures. The comparison clearly shows that there is a difference between the power outputs columns by approximately one order of magnitude (yet with enormous variability). Obviously, this does not allow us to conclude with certainty that Io's total power output—defined as the sum of all simultaneously active hot spots—is ten times higher than previously estimated. However, such a correction might be close to reality for lava lakes only, perhaps overestimated in some cases, considering that Loki Patera, which alone contributes about 10% of the total, is likely already well constrained and therefore should not be revised upward. For comparison, the total estimated global heat flow from Keck Infrared Telescope Facility (IRTF) infrared photometry is around 100 TW (Veeder et al., 1994; 2015), which is in good agreement with the estimation from Io's heat dissipation by Lainey et al., (2009). Veeder et al. (2015) also noted observations of the volcanic centers could only account for ~54% of the total measured heat flow, leaving a large fraction of a missing heat flow source. Similar results for total heat flow were obtained by the Galileo Photopolarimeter Radiometer (PPR) instrument, which could see at longer wavelengths and below 200 K (Rathbun et al., 2004).

However, earlier global estimates based on low-resolution infrared datasets systematically underestimated the contribution of individual paterae. Lava lakes with a hot, continuously renewed perimeter surrounding a warm, crusted interior emit most of their power from narrow regions at very high temperature. When these zones are unresolved, their signal is diluted within mixed pixels, leading to erroneously low brightness temperatures. This bias was further amplified by simplified single-temperature blackbody fits and incomplete spectral coverage, which could not accommodate the true thermal complexity of these systems. High-resolution imaging and multi-component thermal modelling now capture both the hot margins and the cooler interior, yielding power estimates that exceed earlier values by one to two orders of magnitude. Thus, the apparent inconsistency between previously derived global heat flow and newly measured per-feature power simply reflects the limited resolving capability of earlier datasets. The revised, higher power estimates are not contradictory but instead reveal a more realistic heat budget for Io's volcanic system.

We therefore suggest that increasing Io's lava-lake contribution to the global power estimate by up to a factor of ten could be plausible, but only once a comprehensive, high-resolution survey of the thermal output of all paterae becomes available. In the meantime, it must be kept in mind that Io exhibits a wide variety of volcanic manifestations—not all of which are lava lakes—making it practically impossible to provide an accurate and definitive answer to this question.

While these calculations demonstrate beyond any doubt the difficulty of determining the power output for most of Io's paterae—since we lack crustal temperature measurements for many of them—another important aspect must be considered. Mura et al. (2024a) showed that paterae can transition between different morphological and thermal states (with or without the characteristic ring feature). This behavior likely resembles that of Loki Patera (Rathbun et al., 2002).

Even if such transitions are not strictly periodic, JIRAM images clearly show that the same patera (Mura et al., 2024a, Figures 3A and 3B) can display markedly different crustal temperatures. This is a clear indication that the system is not in thermal equilibrium: as the crust cools, it becomes progressively less efficient at radiating the heat coming from Io's interior, effectively acting as a lid over the molten interior. Ultimately, this means that measuring the power output of the system at a single epoch—or even at a few discrete epochs—does not provide a reliable indication of the heat injected into the paterae system from Io's interior, until we achieve a proper understanding of its temporal variability, analogous to what has been observed for Loki.

## 6.2 Crust age and resurfacing

Only a few active lava lakes exist on Earth (Lev et al., 2019), each with characteristics that may inform interpretations of Io's lava lakes. Erta Ale (Ethiopia), Ambrym (Vanuatu) and Nyiragongo (Democratic Republic of Congo) are perhaps the most relevant to Io, although past activity of Kupaianaha (Hawaii) is also relevant (Flynn et al., 1993). On Earth, active lava lakes represent open-vent volcanism in which a body of lava accumulates at the top of the magma column (Lev et al., 2019). The longevity of each lava lake reflects a balance of cooling and outgassing occurring at the surface and input of hot and gas-rich magma from below. Witham and Llewellin (2006) developed a generic model for a lava lake system wherein magmatic pressure at the base of the conduit balances the pressure in the underlying magma reservoir. Using satellite data for the ava lakes at Erebus, Erta Ale and Puʻu ʻOʻo (Hawaiʻi), Harris et al. (1999) concluded that lakes do not necessarily imply endogenous activity but may instead be simply due to recycling of magma in molten reservoirs.

Field observations of the Kupaianaha lava lake in Hawaiʻi between 1987 and 1989 provide insights into lava lake resurfacing. Observations recorded lake levels between 1–20 m below the rim, with an ~50 m diameter lake and a ~30×10 m channel downslope (Flynn et al., 1993). The lake surface moved at ~0.5–1.0 m/s and foundered at the edge above an active lava tube (Kauahikaua et al., 1996). Occasionally, central portions ruptured to expose incandescent lava for tens of seconds, then cooled and solidified. At no time incandescent lava around the perimeter of the entire lake, comparable to the "bright rings" on Io, was observed. Instead, break-ups of the crust were constrained to the segments of the surface as they were being actively subducted above the active lava tube.

However, our estimation of the resurfacing time interval of the Io lava lakes opens some interesting implications for lava lake dynamics. For instance, if our estimation of a resurfacing rate of approximately a decade is correct, then many lava lakes should have experienced three or four such events within the spacecraft era. Voyager 1 and 2 data collected in 1979 should have revealed lava surfaces that had been totally reworked by the time Galileo images were collected between 1996 and 2000. Similarly, Galileo-era images should show lake surfaces that had been destroyed by the time JIRAM collected data between 2023 and 2024. To our knowledge, no decadal changes in lake surface morphology have been reported. Hampering these determinations is the fact that high or even moderate resolution repeat imaging of patera surfaces has not occurred for the vast majority of Io's surface. The best resolution imaging from JunoCam is from the north polar region, which was especially lacking in coverage by Galileo, so studies of change over time with the existing dataset are challenging (Ravine et al. 2025).

Particularly for the larger patera such as Dazhbog (11,000 km$^2$) and Babbar (9,600 km$^2$) the process by which resurfacing takes place is tantalizingly unclear. Did the surface of the entire lava lake founder or overturn at almost the same time, or did segments resurface early and other parts later? It is also unclear whether resurfacing is initiated at the edge of the lake or near the middle of the lake. Resolving this place of onset for lake resurfacing would be valuable in modeling the geometry of the underlying magma chamber. JunoCam images of the surface of the giant Loki and several other nearby paterae at high phase angle reveal extremely flat, reflective surfaces, characteristic of recent lava, while other similarly dark lava flows at low phase do not always show the same reflectivity. This is consistent with relatively recent resurfacing in those patera floors (Ravine et al. 2025). Time-series images with a temporal resolution of a few months over, perhaps, a decade would probably be needed to resolve this issue but the fact that JIRAM data consistently show a uniform crustal thickness (assumed here based on the observed uniform surface temperature, which scales with crust thickness) suggest a relatively quick "complete" resurfacing event at each lava lake. How the age of the lake crust relates to the infrared bright rings is also intriguing. Some form of gravitationally-driven pistoning, as we proposed in the past (Mura et al., 2024a; 2025a), is a possible mechanism for bright ring formation due to the tidal interaction of Io and Jupiter; poroelastic flow has also been proposed (de Kleer et al., 2019). Although crustal plate motion driven by magma convection has also been proposed, this scenario is somewhat less consistent with the observations (Mura et al., 2024a). If the bright rings form by disruption of the crust with the patera walls during this pistoning, then one might expect the most recently formed crusts (e.g., Amaterasu) to display some deformation. As with temporal changes of the lava lakes, this question can only be resolved once future imaging and thermal data are obtained.

Acknowledgments

We thank Agenzia Spaziale Italiana (ASI) for the support of the JIRAM contribution to the Juno mission. This work is funded by the ASI–INAF Addendum n. 2016-23-H.3-2023 to grant 2016-23-H.0. Part of this work was performed at the Jet Propulsion Laboratory, California Institute of Technology, under contract with NASA.

Open Research

The data used in this study comes from Juno and the Jovian Infrared Auroral Mapper (JIRAM) and is publicly available on the Planetary Database System in PDS4 format (Adriani et al., 2019). [The dataset is also available from ](#)Mura, 2024. The SPICE bundles used to calibrate the viewing geometry are also publicly available in the Planetary Database System via the Navigation and Ancillary Information Facility (NAIF). The Voyager-Galileo Io data used for our analysis is available through Williams et al., (2011a) on the U.S. Geological Survey Scientific Investigations website (Williams et al., 2011b).

Corresponding author

Correspondence to Alessandro Mura (alessandro.mura@inaf.it)

Competing interests

The authors declare no competing interests.

# 8 Appendix

## 8.1 Estimation of crust temperature and total power

The total power in the M band is simply calculated by integrating the JIRAM signal within a box containing the lava lake. In Mura (2025), we provide all the images used in this work; the box we refer to is the largest one shown in the figures. The median signal outside this box is used to estimate the background level, which is subtracted from the measured radiance as an initial step.

As for the estimation of the radiance on the crust, in order to avoid overestimating it and to obtain a more robust result, we follow three different approaches. In cases where a sufficiently large crustal region is visible within the patera, we define a box that contains the entire crust but does not include portions outside the ring (the intermediate-sized box in the figures S1 to S36 in Mura, 2026). We then compute the kernel density estimation (KDE) of the radiance (not to be confused with the ks-density of the temperature shown in Figure 6) within this box and take the lowest-value peak $R_p$, which clearly represents the group of pixels with the lowest radiance. It should be noted that as long as this box includes the crust of the patera but does not extend beyond the hot spot, the result is not sensitive to the specific choice of the box, because the first KDE peak remains unchanged, being essentially the cluster of data with the lowest radiance, i.e., the crust.

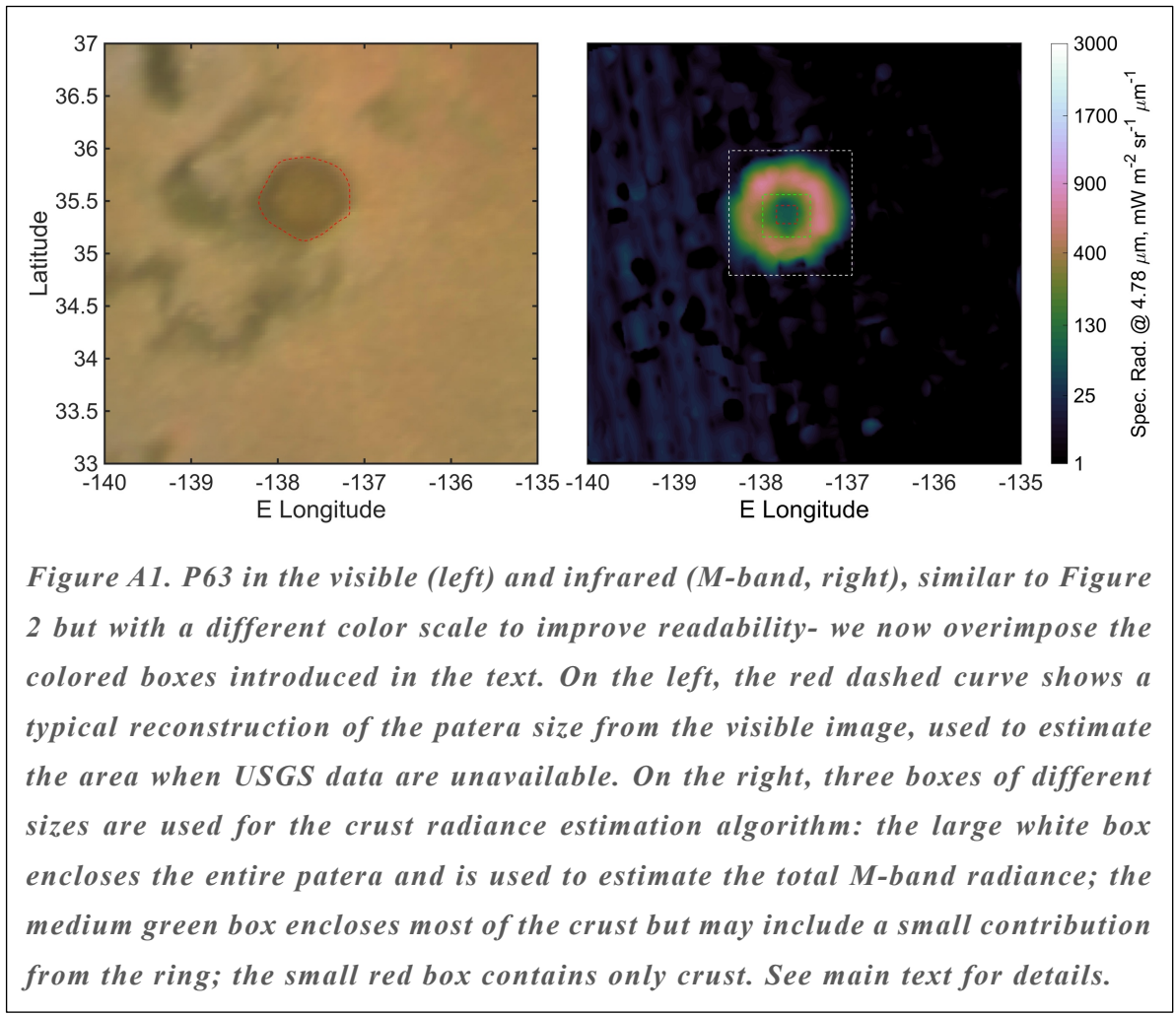

*Figure A1. P63 in the visible (left) and infrared (M-band, right), similar to Figure 2 but with a different color scale to improve readability- we now overimpose the colored boxes introduced in the text. On the left, the red dashed curve shows a typical reconstruction of the patera size from the visible image, used to estimate the area when USGS data are unavailable. On the right, three boxes of different sizes are used for the crust radiance estimation algorithm: the large white box encloses the entire patera and is used to estimate the total M-band radiance; the medium green box encloses most of the crust but may include a small contribution from the ring; the small red box contains only crust. See main text for details.*

Subsequently, we select a smaller box (also shown in the figures S1 to S36 in Mura, 2026) that contains only regions with uniform radiance within the ring and compute its median $R_m$. The two values, $R_p$ and $R_m$, are compared and validated when they are reasonably similar (a difference of 1 or 2 K); in such cases, the result is accepted. When $R_m$ is significantly lower than $R_p$, this more conservative value is used instead, in order to avoid overestimating the crust temperature.

Finally, in cases where the spatial resolution does not allow identifying a sufficiently extended crustal region, we take the minimum radiance value within the ring. These cases are labelled as "uncertain," to indicate that their estimation is less robust compared to the others, and their values are not used in the computation of average quantities derived from $R$.

The uncertainty on this method can be estimated by evaluating the FWHM of the KDE peak $R_p$, or, if method 2 is preferred (see table 2 for these cases), the standard deviation of the distribution of radiances of which $R_m$ is the median.

The emissivity is assumed as in Mura et al. (2024a), equal to 0.95 below 500 K and decreasing to 0.6 at T = 1500 K (in practice, the emissivity for the temperatures considered here is almost always close to 1). From the radiance and emissivity, we derive the brightness temperature. The fundamental assumption for this estimate is that brightness temperature is meaningful only when there are no sub-pixel structures; therefore, at the spatial resolution available here, it provides a good estimate of the crust temperature.

Knowing the temperature, we compute the blackbody spectrum, again accounting for emissivity, integrate it, and multiply it by the area of the patera. The crust area practically coincides with the area of the patera itself, since the "hot" ring has negligible thickness compared to the lava lake diameter. The dimensions of the paterae are estimated by measuring the area within the red dashed region in the visible-wavelength image (right panel of figures S1 to S36 in Mura, 2026); as a check, we compare this area with the values reported in the USGS database (for example, for Chors: https://planetarynames.wr.usgs.gov/Feature/1198). In all examined cases, our estimates are never larger than the USGS (United States Geological

Survey) values, confirming that we are not overestimating the area and therefore the total power. In a few cases, since there was no available image in the visible range, the dimension has been estimated from the IR map.

Once the area, temperature, and M-band emission of the ring are known, it is straightforward to compute the total power of the crust as described in the example in Section 3. In cases where there is a clear (though minor) contribution from the hotter ring, it can optionally be added to the total. Since the ratio between total power and M-band power never drops below ~13 (Tosi et al., 2025), we can conservatively use this factor to scale the high-temperature component of the hot spot radiance.

Uncertainties on the crust temperature (on average, ~5%) are derived from the uncertainties on $R_m$ or $R_p$, same is for the uncertainties on the total power (on average, ~10%). We prefer giving an average value rather than the single ones because, for some cases, the procedure gives an unrealistically low uncertainty for the temperature (i.e. 1 K); the average uncertainty, instead, is more representative of the intrinsic uncertainty that the method brings.

## 8.2 Estimation of resurfacing time scale

Once the crust temperature is obtained, we infer the crust age by simply interpolating the temperature with the model of crust cooling in Mura et al., (2025b), which is basically the same as the one by de Kleer and de Pater (2017). Then, in order to estimate the characteristic timescale of lava lake crust overturns on Io, we assume that overturn events follow a Poisson process with mean recurrence time $\tau$. In this framework, the waiting times between events are exponentially distributed with probability density function:

$$f(t) = \frac{1}{\tau} e^{-t/\tau}$$

Our dataset consists of $N$ lava lakes, for each of which we measure the crust age $t_i$, defined as the elapsed time since the last overturn at the moment of observation (or, more precisely, the inferred crust age, from the crust temperature). Owing to the memoryless property of the exponential distribution, these elapsed times are themselves independent realizations of the same exponential law. The likelihood function for $\tau$ is therefore:

$$L(\tau) = \prod_{i=1}^{N} f(t_i|\tau) = \prod_{i=1}^{N} \frac{1}{\tau} e^{-t_i/\tau}$$

or:

$$L(\tau) = \frac{1}{\tau^N} e^{-\sum_{i=1}^{N} t_i/\tau}$$

with corresponding log-likelihood:

$$\ell(\tau) = -N \ln(\tau) - \frac{1}{\tau} \sum_{i=1}^{N} t_i$$

Maximization of the log-likelihood is straightforward, since its derivative of vanishes at:

$$\hat{\tau} = \frac{1}{N} \sum_{i=1}^{N} t_i$$

which yields the maximum likelihood estimator, that is the arithmetic mean of the observed crust ages.

In some cases, not all the crust ages are directly measured. For the majority $M$ of the lakes, $t_i$ is measured, but for the others, only a lower limit is available (e.g., $t_i > c$). In this situation, the likelihood must be modified to account for the right-censored nature of these data. For each censored observation, the probability of obtaining $t_i > c$ under the exponential model is:

$$P(t_i > c | \tau) = \int_{c}^{\infty} \frac{1}{\tau} e^{-t/\tau} dt = e^{-c/\tau}$$

Therefore, for a dataset containing $M$ measured and $N-M$ censored values, the likelihood becomes:

$$L(\tau) = \frac{1}{\tau^M} e^{-\sum_{i=1}^{M} t_i/\tau - \sum_{i=M}^{N} c/\tau}$$

The rest of the derivation follows analogously, with the censored limits contributing to the likelihood through the exponential term only.

$$\hat{\tau} = \frac{1}{M} \left( \sum_{i=1}^{M} t_i + (N-M) \cdot c \right)$$

In words: censored observations enter only through their censoring time via the survival function, and they increase the total time in the numerator while reducing the effective sample size to the count of uncensored data.

For completeness, we also consider a different scenario in which each lava lake overturns in a regular, periodic fashion, rather than stochastically. This would correspond to a "Loki Patera–like" behaviour, where a given lake follows a deterministic cycle of duration $T_i$. In such a case, the observed crust ages $t_i$ are no longer drawn from an exponential distribution, but rather represent random phases within each individual cycle. The expected value of the measured crust age for lake $i$ is then $T_i/2$, as the observation is equally likely to occur at any point in the cycle. Consequently, the arithmetic mean of the measured ages across $N$ lakes would converge to half of the average cycle length ($\bar{T}$). Thus, under this non-Poissonian and strictly periodic assumption, the mean of the observed crust ages would not yield the characteristic timescale of overturns directly, but rather a biased estimator corresponding to one half of the true average period. Note that, under this assumption, the above formula for the case of right-censored data gives a lower limit of $\bar{T}/2$.

Finally, the purely speculative hypothesis—that we are observing resurfacing events all with the same periodicity, which has no physical basis—can be tested using a Kolmogorov-Smirnov test against a uniform distribution with an arbitrarily chosen upper limit greater than 10 years. Performing this test with upper limits of 15 or 20 years (and even more so for values above 20) yields extremely low p-values, clearly rejecting the hypothesis.